\documentclass[sigconf]{acmart}
\usepackage{hyperref}

\setcopyright{none}

\acmDOI{}

\acmISBN{}

\acmConference[Preprint (2023)]{}{IDE}{UiS}
\acmBooktitle{}
\copyrightyear{2023}

\newcommand{\supervisors}[1]{\thanks{Permission to make digital or hard copies of all or part of this work for personal or classroom use is granted without fee provided that copies are not made or distributed for profit or commercial advantage and that copies bear this notice and the full citation on the first page. Copyrights for components of this work owned by others than the author(s) must be honored. Abstracting with credit is permitted.  To copy otherwise, or republish, to post on servers or to  redistribute to lists, requires prior specific permission and\hspace*{.5pt}/or  a fee.}}

\usepackage{xspace}

\usepackage{tikz}
\usepackage{pgfplots}
\pgfplotsset{compat=newest}

\title{Digital and Cloud Forensic Challenges}
\begin{document}
\hypersetup{
    colorlinks=true,
    linkcolor=blue,
    filecolor=red,      
    urlcolor=red,
    citecolor=blue,
}

\author{Ahmed MohanRaj Alenezi }
\affiliation{Independent Forensics Researcher, Auckland, New Zealand}
\email{amam88820@gmail.com}

\supervisors{Leander Jehl and Hein Meling}

\begin{abstract}
\textbf{Digital forensics and cloud forensics are increasingly important fields that face a range of challenges. This study aims to assess the general challenges faced in these fields. A literature review was conducted to identify the major challenges in digital and cloud forensics, including data acquisition, data analysis, data preservation, privacy concerns, and legal issues. The challenges were analyzed in detail, considering the reasons why they are challenges, the impact they have on digital and cloud forensics, and any potential solutions. The study concludes that the challenges faced in digital and cloud forensics are significant and varied, and that addressing these challenges is critical for the effective and efficient use of digital and cloud forensics in investigations. This study provides a valuable overview of the current state of digital and cloud forensic challenges and can help guide future research in this important field.
}

\end{abstract}


\maketitle


\section{Introduction}
\label{sec:introduction}

\noindent
Digital and cloud forensics \cite{garfinkel2010digital} are increasingly critical fields in today's digital age. As the use of digital devices and cloud services continues to grow, the need for effective and efficient digital \cite{aarnes2017digital,casey2009handbook} and cloud forensic investigations has become more important than ever However, with the increasing complexity of digital devices \cite{kebande2016generic,kebande2020quantifying} and cloud services, digital and cloud forensic investigations have become more challenging. This paper aims to assess the challenges faced in digital and cloud forensics, with a focus on data acquisition, data analysis, data preservation, privacy concerns, and legal issues.

As technology continues to advance, the challenges facing digital forensics \cite{yassin2020cloud,taylor2011forensic} and cloud forensics have become more complex. In particular, the widespread use of cloud computing has presented a range of challenges for forensic investigators, including issues related to data acquisition, preservation, and analysis. Despite these challenges, the ability to collect and analyze digital and cloud-based evidence has become increasingly important in the investigation of cybercrime, terrorism, and other forms of criminal activity.

The challenges of digital and cloud forensics   \cite{ruan2011cloud,birk2011technical,orton2013legal} \cite{chung2012digital,farina2015overview,kebande2018novel} can be significant and varied. Data acquisition, for example, can be challenging due to the wide range of digital devices and cloud services available, as well as the large volumes of data that need to be collected and analyzed. Data analysis can also be challenging, as the sheer amount of data that needs to be processed can make it difficult to identify relevant information. Data preservation is another significant challenge, as digital data can be easily altered or deleted if not properly preserved \cite{damshenas2012forensics,guo2012forensic,reilly2011cloud}. There need to be some guidelines for conducting these processes to alleviate these forensic challenges in the cloud \cite{achar2022cloud,teing2017forensic,thethi2014digital} \cite{owen2011analysis},\cite{alex2017forensics,kebande2018novel}.

Privacy concerns are also a major challenge in digital and cloud forensics. Digital devices and cloud services often contain sensitive personal information, and investigators must take care to ensure that this information is properly protected. Legal issues can also pose significant challenges in digital and cloud forensics, as investigators must navigate complex legal frameworks and adhere to strict rules and regulations.

In this paper, we will assess these challenges in detail, considering the reasons why they are challenges, the impact they have on digital and cloud forensics, and any potential solutions. By understanding these challenges, we can better address them and improve the effectiveness and efficiency of digital and cloud forensic investigations.

The remainder of the paper is organized as follows: Section 2 gives the background to the study, while Section 3 discusses related works and the significance of the current research work. Section 4 presents the Digital forensics and Section 5 presents cloud forensics challenges. results and Discussion are given in Section 6 and a Conclusion in Section 7.

\begin{figure*}[h]
   \centering
   \includegraphics[width=0.5\textwidth]{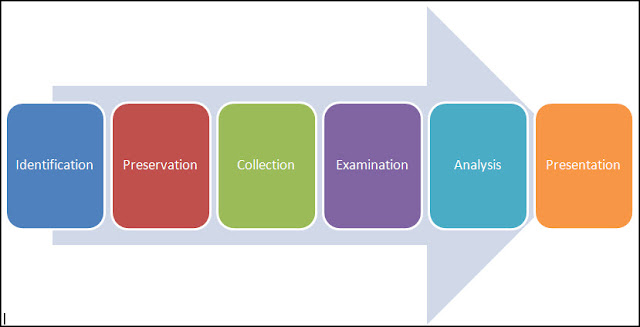}
    \caption{Digital Forensic Process}
    \label{fig:example}
\end{figure*}

\section{Background}
\label{sec:background}
To understand the challenges faced in digital and cloud forensics, it is important to have a clear understanding of the background of these fields and the factors that have contributed to their growth and importance in today's society.. This section gives a background on As a result, this section gives background on digital and cloud forensics, providing an overview of the history and development of these fields, as well as the key factors that have contributed to their growth and importance. The section will also examine the role of digital and cloud forensics in modern society and highlight some of the key challenges faced in these fields. Overall, this background section will provide a foundation for understanding the significance of digital and cloud forensics, and how these fields have evolved over time to become critical components of modern investigations \cite{aarnes2017digital,casey2009handbook}.

\subsection{Digital Forensics}
Digital forensics is a field that involves the collection, preservation, and analysis of digital data for use as evidence in investigations. Digital forensics has become increasingly important in recent years, as the use of digital devices has become ubiquitous in today's society. Digital devices, including computers, smartphones, and tablets, are used to store a wide range of data, including personal and financial information, communications, and other sensitive data \cite{pollitt2010history,vincze2016challenges,richard2006next,kebande2016requirements,delp2009digital,kavrestad2020fundamentals}. This data can be critical in investigations related to cybercrime, terrorism, and other forms of criminal activity \cite{ieong2006forza,adedayo2016big,roussev2009hashing,montasari2017standardised,zawoad2015digital,quick2016big}.

The history of digital forensics can be traced back to the 1980s when computer forensics was first used by law enforcement agencies in investigations \cite{pollitt2010history,casey2019chequered,raghavan2013digital,rogers2017technology,sachowski2018digital}. At that time, computer forensics primarily involved the collection and analysis of data stored on computer hard drives. Over time, however, digital forensics has evolved to include the analysis of data stored on a wide range of digital devices, including smartphones, tablets, and other mobile devices \cite{yates2010practical,kebande2015obfuscating,rogers2017technology,al2020review,gentry2019seaker,dogan2017analysis}.

Digital forensics involves several key steps, including data collection, preservation, analysis, and reporting as is shown in Figure \ref{fig:example}. Data collection involves the identification and acquisition of relevant digital data, while data preservation involves the proper storage and protection of this data to ensure its integrity. Data analysis involves the processing and examination of digital data to identify relevant information, while reporting involves presenting the findings of the analysis in a clear and concise manner \cite{flaglien2017digital,reith2002examination,kebande2018novel,kebande2014cloud,varol2017review,alharbi2011proactive,agarwal2011systematic,kebande2015towards}.

Digital forensics has become a critical component of many investigations, including those related to cybercrime, terrorism, and corporate fraud. In many cases, digital data is the only source of evidence available, making the ability to collect, analyze, and preserve this data critical for successful investigations \cite{holt2022cybercrime,harbawi2016role,gladyshev2015digital,muhammad2022role,sikos2021ai}. Digital forensics is also essential in cases where traditional investigative techniques are not effective, such as in cases involving encrypted data or data stored on cloud services \cite{atlam2020security,makura2021digital,wiles2011best,chawki2015cybercrime,prayudi2015proposed,almulla2014state,simou2016survey,simou2016survey}.

\subsection{Cloud Forensics }

Cloud computing has become a ubiquitous technology in recent years, providing businesses and individuals with a range of benefits, including scalability, cost savings, and improved collaboration. Cloud computing involves the use of remote servers to store, manage, and process data, which can be accessed from anywhere with an internet connection. However, as with any technology, the use of cloud computing also presents several challenges, particularly when it comes to digital forensics.

Cloud forensics is a branch of digital forensics that focuses on the investigation of data stored on cloud servers \cite{simou2016survey,zargari2012cloud,almulla2013cloud,alqahtany2015cloud,kebande2018digital,kebande2020ontology,karie2019diverging}. The use of cloud services presents several unique challenges for forensic investigators, including issues related to data acquisition, preservation, and analysis. One of the key challenges of cloud forensics is the fact that data is often stored on servers located in multiple geographic locations, which can make it difficult to identify and collect relevant data. Additionally, cloud providers may have different policies regarding data retention and deletion, which can make it difficult to preserve data for investigative purposes \cite{ruan2013cloud,pichan2015cloud,kebande2015functional,kebande2015obfuscating,kebande2019comparative,
zawoad2013cloud,almulla2014state,zargari2012cloud,almulla2013cloud,alqahtany2015cloud}.

Another challenge of cloud forensics is the fact that cloud servers may be shared among multiple users, which can make it difficult to identify and isolate relevant data. Cloud providers may also use encryption and other security measures to protect data, which can make it difficult to access and analyze data for investigative purposes. Finally, the dynamic nature of cloud computing, which involves the rapid deployment and scaling of resources, can make it difficult to track and preserve data over time.

Despite these challenges, cloud forensics has become an increasingly important field in recent years, particularly as the use of cloud services has become more widespread. In many cases, cloud data may be the only source of evidence available in investigations, making the ability to collect, analyze, and preserve this data critical for successful investigations. As such, cloud forensics has become an essential tool for law enforcement agencies, cybersecurity professionals, and other investigators who are tasked with investigating cybercrime, terrorism, and other forms of criminal activity.

\section{Related Work}
\label{sec:RW}
There has been a significant amount of research conducted in the field of digital forensics and cloud forensics, with a particular focus on the challenges associated with data acquisition, preservation, and analysis. Researchers have developed a range of tools and techniques designed to help forensic investigators overcome these challenges and successfully recover and preserve digital and cloud-based evidence.

In the field of digital forensics, researchers have developed a variety of tools and techniques for data acquisition, including imaging tools that can create a bit-by-bit copy of a digital device, as well as tools for extracting data from a live system. Researchers have also developed techniques for analyzing digital evidence, including keyword searching, timeline analysis, and link analysis.

In the field of cloud forensics, researchers have focused on the challenges associated with data acquisition and preservation in cloud-based environments. In particular, researchers have developed tools and techniques for acquiring data from cloud servers, including methods for acquiring data from cloud storage providers such as Dropbox and Google Drive. Researchers have also explored techniques for preserving data in the cloud, including techniques for preserving metadata and ensuring data integrity.

\section{CHALLENGES IN DIGITAL FORENSICS}
\label{sec:Challenges}
The effective investigation and prosecution of digital crimes require overcoming several technical, legal, and ethical challenges. Digital forensic investigators face a dynamic and complex environment that continually evolves and requires up-to-date knowledge, tools, and expertise. Identifying and addressing these challenges is critical to the success of digital forensic investigations, and the outcomes of legal proceedings that rely on digital evidence. In this section, we will examine some of the most significant challenges faced by digital forensic investigators. Significant challenges have been identified in relevant research that has a focus on the technical challenges \cite{mohay2005technical,kebande2015adding,kebande2019cfraas,
james2013challenges,
strandberg2022systematic,
montasari2019next,
kebande2018digital,
pichan2015cloud}.

Technical digital forensics challenges refer to the difficulties encountered in the technical aspects of digital forensics investigations. Some of these challenges include the rapidly changing technology landscape, with new devices, operating systems, and applications being developed constantly, making it difficult for digital forensic investigators to keep up  \cite{chen2014cloud,kebande2019comparative,kebande2022industrial,losavio2018internet,casino2022research,birk2011technical,stoyanova2020survey,
mohammmed2020survey}. Additionally, the use of encryption and other security measures to protect data can make it challenging to access and analyze digital evidence. Another challenge is the large volumes of data that must be analyzed, which can be time-consuming and resource-intensive. Finally, the complexity of digital systems and networks can make it difficult to identify the source of security incidents or data breaches, as well as the extent of the damage done. Overall, these technical challenges highlight the need for ongoing research and development in digital forensics tools and techniques \cite{mouhtaropoulos2014digital,kebande2020mapping,
simou2014cloud,
mantas2022watches,
kebande2020holistic,
arafat2017technical,
ricci2019blockchain,
alenezi2019experts}. These are shown in summary in  Table 1 and 2 respectively.
\begin{table*}[h]
\centering
\caption{Technical challenges in digital forensics}
\label{tab:challenges}
\begin{tabular}{|p{4cm}|p{9cm}|}
\hline
\textbf{Technical Challenge} & \textbf{Description} \\\hline
Data volume & Overwhelming amount of digital data to analyze in investigations \\\hline
Data complexity & Digital data can be complex, structured and unstructured, metadata, system logs, hidden files \\\hline
Data storage & Data stored in different locations, local and remote servers, cloud storage, physical devices \\\hline
Data encryption & Use of encryption to protect data from unauthorized access \\\hline
Data spoliation & Data can be easily modified, deleted, or overwritten, intentionally or accidentally \\\hline
Emerging technologies & Rapid pace of technological innovation means investigators must constantly adapt \\\hline
Cross-jurisdictional challenges & Investigations often involve multiple jurisdictions, creating challenges in information sharing and coordination \\\hline
\end{tabular}
\end{table*}

\subsection{Technical Challenges}

\begin{itemize}
    \item Data volume: The sheer amount of digital data that must be analyzed in digital forensic investigations can be overwhelming, particularly in cases where multiple devices are involved. The sheer volume of data can be overwhelming for investigators, leading to longer investigation times and increased costs. Additionally, as the amount of data grows, the likelihood of finding relevant evidence may decrease, as investigators must sift through large amounts of irrelevant data to find the crucial pieces of evidence needed for the investigation. Therefore, managing and analyzing large volumes of data in a timely and efficient manner is a critical challenge in digital forensics. This challenge requires the use of powerful tools and techniques for data analysis and processing, as well as strategies for managing the volume of data to focus on relevant information\cite{sree2020data,makura2020proactive,kebandeadding,karie2016generic}.
    \item Data complexity: Digital data is often complex and can include structured and unstructured data, metadata, system logs, and hidden or deleted files, which can make analysis difficult and time-consuming. Digital data is often complex and can include a wide variety of different types of data, such as structured and unstructured data, metadata, system logs, and hidden or deleted files. This can make analysis difficult and time-consuming, as investigators must carefully examine each piece of data to determine its relevance to the investigation. In addition, the increasing use of encryption and other security measures to protect data can make it even more challenging to extract meaningful information from digital devices. As a result, digital forensic investigators must be highly skilled in a variety of technical areas, including data analysis, cryptography, and information security, in order to effectively analyze and interpret complex digital data\cite{baror2021framework,baror2022conceptual,kavrestad2020fundamentals,al2020categorization}.
    \item Data storage: Digital data can be stored in a variety of different locations, including local and remote servers, cloud storage, and physical devices. The challenge for investigators is to identify and collect all relevant data from these disparate sources. Digital data can be stored in a variety of different locations, including local and remote servers, cloud storage, and physical devices. This can create significant challenges for digital forensic investigators, as they must identify and collect all relevant data from these disparate sources. In addition, the storage medium itself can present challenges, as some devices may be damaged or physically inaccessible, while others may require specialized tools or techniques to access. As a result, digital forensic investigators must be well-versed in a variety of different technologies and storage systems, and must be able to adapt their techniques to effectively collect and analyze data from any storage medium\cite{kebande2022finite,agarwal2011systematic,zawali2021realising,al2020review}.
    \item Data encryption: The use of encryption to protect data from unauthorized access can make it difficult for investigators to access and analyze digital data. The use of encryption to protect data from unauthorized access can be a significant challenge for digital forensic investigators. Encrypted data cannot be read or analyzed without the proper decryption key or password, which can be difficult or impossible to obtain. As a result, investigators must employ a variety of different techniques and tools to attempt to break encryption and access encrypted data. This can include brute-force attacks, dictionary attacks, and other methods designed to bypass encryption and gain access to the underlying data\cite{roussev2009hashing,azhan2022error,
kigwana2017proposed,
kigwana2017towards,
hungwe2019scenario}.
    \item Data spoliation: Digital data can be easily modified, deleted, or overwritten, either intentionally or accidentally, which can compromise the integrity of the data and the results of an investigation. Digital data can be easily modified, deleted, or overwritten, either intentionally or accidentally. This can compromise the integrity of the data and the results of an investigation. As a result, digital forensic investigators must take steps to ensure the preservation of digital data throughout the investigation process. This can include making forensic copies of all relevant data, using write-blockers to prevent accidental modification, and implementing strict chain-of-custody procedures to ensure the integrity of the data is maintained\cite{al2020towards}.
    \item Emerging technologies: The rapid pace of technological innovation means that digital forensic investigators must constantly adapt and evolve to stay up-to-date with new devices, operating systems, and applications. The rapid pace of technological innovation means that digital forensic investigators must constantly adapt and evolve to stay up-to-date with new devices, operating systems, and applications. This requires a deep understanding of the latest technologies, as well as the ability to quickly learn and implement new techniques and tools. In addition, investigators must be able to anticipate and identify emerging threats and vulnerabilities, and proactively develop strategies to address these risks.
    \item Cross-jurisdictional challenges: Digital forensic investigations often involve multiple jurisdictions, each with their own legal and regulatory frameworks, which can create challenges in sharing information and coordinating investigations. This can create significant challenges in terms of sharing information and coordinating investigations. Investigators may need to navigate complex legal issues related to data privacy, data protection, and intellectual property, as well\cite{kebande2022finite,kebande2020ontology,kebande2015functional,kebande2020mapping,kebande2021review}.
\end{itemize}

\subsection{Legal Challenges }

Legal challenges in digital forensics arise from the legal and regulatory frameworks governing digital evidence collection, analysis, and presentation in court. These challenges can have a significant impact on the admissibility and reliability of digital evidence and the outcome of legal proceedings. The following are some of the key legal challenges in digital forensics:

\begin{itemize}
\item Cross-border jurisdiction: Digital forensic investigations can span multiple jurisdictions, each with their own legal and regulatory frameworks, which can create challenges in terms of obtaining evidence and sharing information. Digital forensics investigations may span multiple jurisdictions with varying legal frameworks, making it difficult to collect and analyze digital evidence. The lack of harmonized legal frameworks between jurisdictions can also complicate the exchange of information and data. Therefore, investigators must be aware of legal constraints when conducting cross-border investigations and ensure that they comply with local laws.
\item Admissibility of evidence: The admissibility of digital evidence in court can be challenged due to issues such as authenticity, reliability, and chain of custody. The admissibility of digital evidence in court can be challenging due to issues such as authenticity, accuracy, and reliability. The evidence must be collected, stored, and analyzed in accordance with established protocols and must be presented in a manner that is admissible in court\cite{lillquist2002comment,
giannelli2003admissibility,
chaski2012best,
moenssens1983admissibility,
holobinko2012forensic}. Courts generally require that evidence be relevant, reliable, and obtained through legal means. Digital evidence can be challenging to meet these standards due to the potential for data spoliation, the use of encryption, and the need for specialized expertise to collect and analyze digital evidence.
\item Privacy and data protection: Digital forensic investigations can involve the collection and analysis of personal data, which raises concerns about privacy and data protection laws. Privacy and data protection laws, such as the EU General Data Protection Regulation (GDPR), have made it more challenging for investigators to access personal data. These laws impose strict limitations on the processing of personal data, which can hinder digital forensic investigations\cite{law2011protecting,sim2022argus,joshi2019security,nieto2018iot,halboob2015privacy}.
\item Lack of standardization: There is a lack of standardization in digital forensic procedures, which can create inconsistencies in the collection, analysis, and presentation of evidence. One of the key legal challenges in digital forensics is the lack of standardization in legal and regulatory frameworks across different jurisdictions. This can make it difficult to determine which laws and regulations apply to a particular investigation, and can also make it difficult to share information and evidence between jurisdictions. In some cases, different jurisdictions may have conflicting laws and regulations, which can further complicate investigations\cite{ISO/IEC2700,ISO/IEC27043}.

Additionally, there is often a lack of standardization in digital forensic procedures and techniques. Different investigators and organizations may use different tools and methods for analyzing digital evidence, which can lead to inconsistencies and inaccuracies in the results. This lack of standardization can also make it difficult for investigators to share their findings and conclusions with others, particularly if they are using different terminology or procedures.

\item Ethical considerations: Digital forensic investigators must also consider ethical considerations when conducting investigations, such as respecting the privacy and dignity of the individuals whose data is being analyzed. Investigators must ensure that their actions are legal, ethical, and in compliance with industry standards and best practices.
\end{itemize}

\begin{table*}[h]
\centering
\caption{Legal Challenges in Digital Forensics}
\begin{tabular}{|p{4cm}|p{9cm}|}
\hline
\textbf{Legal Challenge} & \textbf{Description}   \\\hline
Cross-border jurisdiction & Digital forensic investigations can span multiple jurisdictions, each with their own legal and regulatory frameworks, which can create challenges in terms of obtaining evidence and sharing information.  \\\hline
Admissibility of evidence & The admissibility of digital evidence in court can be challenged due to issues such as authenticity, reliability, and chain of custody.  \\\hline
Privacy and data protection & Digital forensic investigations can involve the collection and analysis of personal data, which raises concerns about privacy and data protection laws.  \\\hline
Lack of standardization & There is a lack of standardization in digital forensic procedures, which can create inconsistencies in the collection, analysis, and presentation of evidence.  \\\hline
Ethical considerations & Digital forensic investigators must also consider ethical considerations when conducting investigations, such as respecting the privacy and dignity of the individuals whose data is being analyzed as is shown in Table 2.  \\\hline
\end{tabular}

\label{tab:legal_challenges}
\end{table*}

\section{Challenges in Cloud Forensics}
\label{sec:Cloud}
Cloud forensic refers to the process of collecting, analyzing, and preserving digital evidence from cloud-based systems and services. As more and more organizations are shifting their data and services to cloud environments, the demand for cloud forensic is increasing rapidly. However, the nature of cloud environments poses unique challenges to forensic investigators. One of the main challenges is the lack of control over the cloud infrastructure. Unlike traditional computer systems, cloud environments are owned and managed by cloud service providers (CSPs), making it challenging for investigators to access the evidence they need.

Another significant challenge in cloud forensic is the dynamic nature of cloud environments. Cloud systems are highly distributed and complex, with data and applications residing in multiple locations and accessed by various users and devices. This creates a massive volume of data, making it challenging to identify and isolate relevant evidence. Additionally, the data in cloud environments is often encrypted or obfuscated, adding another layer of complexity to forensic investigations.

Cloud forensic investigators also face challenges related to the legal and regulatory landscape. The jurisdictional issues that arise in traditional digital forensics are further complicated in cloud environments, as data can be stored and accessed from multiple locations across the globe. Furthermore, cloud data is often subject to multiple laws and regulations, making it challenging to navigate the legal landscape and obtain the necessary permissions to access and analyze the evidence.

Moreover, the distributed nature of cloud environments means that data may be replicated across multiple servers and data centers, making it challenging to determine the location of relevant data for investigation purposes. Furthermore, encryption and access controls may also limit access to data and require additional authentication mechanisms for investigators to gain access to the data.

Cloud forensic readiness refers to the preparedness of an organization to effectively and efficiently conduct digital forensic investigations on data stored in the cloud. It involves having appropriate policies, procedures, and technical controls in place to ensure that digital evidence can be collected, preserved, and analyzed in a manner that is consistent with legal and regulatory requirements.

Lastly, the lack of standardized cloud forensics procedures and tools creates another challenge. There is a need for standardized procedures and tools for cloud forensics investigations, which can help to ensure the consistency and reliability of the results. The lack of such standardization can lead to variations in the quality and admissibility of the evidence presented in court.

\begin{itemize}
\item Lack of understanding: Many organizations lack a clear understanding of what cloud forensic readiness entails, including how to properly configure their cloud environments and how to preserve evidence in the cloud.

\item Lack of standardization: The lack of standardization in cloud environments can create challenges in terms of preserving evidence and ensuring its admissibility in court. Different cloud providers have different methods for preserving evidence, and these methods may not always be compatible with forensic investigations.

\item Complex architectures: Cloud environments can be complex, with multiple layers of infrastructure and numerous interconnected services. This complexity can make it difficult to determine which data and systems are relevant to a digital forensic investigation.

\item Dynamic nature of cloud environments: Cloud environments are dynamic, with resources constantly being provisioned and de-provisioned. This can make it challenging to identify the source of a security incident or data breach.
\item Legal and regulatory issues: The legal and regulatory issues associated with cloud computing, such as cross-border data transfer, can create additional challenges in terms of preserving evidence and ensuring its admissibility in court.
\item Cost considerations: Preparing a cloud environment for forensic investigations can be expensive. E.g if an organization needs to implement specialized tools or hire external consultants to assist with the process.
\end{itemize}

\section{Result and Discussion}
\label {sec:result}
Digital and cloud forensics face various challenges that can affect the accuracy and efficiency of the forensic investigations. One of the main challenges in digital forensics is the cross-border jurisdiction, where investigations can span multiple jurisdictions, each with its legal and regulatory frameworks. This can create challenges in terms of obtaining evidence and sharing information. Additionally, admissibility of digital evidence in court can be challenged due to issues such as authenticity, reliability, and chain of custody. Privacy and data protection also pose a challenge since digital forensic investigations involve the collection and analysis of personal data, raising concerns about privacy and data protection laws\cite{kebande2018forensic}.

Cloud forensics, on the other hand, faces unique challenges compared to traditional digital forensics. The lack of standardization in cloud environments can create challenges in preserving evidence and ensuring its admissibility in court. Moreover, cloud environments can be complex, making it difficult to determine which data and systems are relevant to a digital forensic investigation. The dynamic nature of cloud environments, where resources are continually being provisioned and de-provisioned, also presents challenges in identifying the source of a security incident or data breach. Legal and regulatory issues associated with cloud computing, such as cross-border data transfer, can create additional challenges in preserving evidence and ensuring its admissibility in court. Finally, preparing a cloud environment for forensic investigations can be expensive, particularly if an organization needs to implement specialized tools or hire external consultants to assist with the process\cite{karie2018knowledge,kebande2018towards}.

The challenges faced by digital and cloud forensics are multifaceted and complex. In terms of digital forensics, the challenges include cross-border jurisdiction, admissibility of evidence, privacy and data protection, lack of standardization, and ethical considerations. Cloud forensic readiness, on the other hand, is challenged by a lack of understanding, lack of standardization, complex architectures, the dynamic nature of cloud environments, legal and regulatory issues, and cost considerations.

These challenges can hinder digital and cloud forensic investigations and impact the admissibility of evidence in court. The lack of standardization in both digital and cloud environments can lead to inconsistencies in evidence collection, analysis, and presentation, which can further impede investigations. The dynamic nature of cloud environments can make it difficult to preserve evidence, and the legal and regulatory issues associated with cloud computing can create additional challenges in terms of preserving evidence and ensuring its admissibility in court.

The challenges faced in digital and cloud forensics emphasize the need for standardization, collaboration, and training of forensic professionals. Standardization of procedures, policies, and guidelines in digital and cloud forensics can help in ensuring consistency in the collection, analysis, and presentation of evidence. Collaboration among stakeholders such as cloud service providers, forensic professionals, and legal professionals can also help in addressing the unique challenges presented by cloud computing. Additionally, training of forensic professionals and other stakeholders can help in improving their understanding of the challenges and the best practices in digital and cloud forensics.

\section{Conclusion and Future Work}
\label{sec:conclude}
Digital and cloud forensics are critical in today's world where digital technology plays a significant role in various aspects of our lives. This paper has identified and discussed various challenges that digital and cloud forensics face. These challenges include cross-border jurisdiction, admissibility of evidence, privacy and data protection, lack of standardization, complex architectures, dynamic nature of cloud environments, legal and regulatory issues, and cost considerations.

The challenges identified in this study have implications for the practice of digital and cloud forensics. Addressing these challenges will require the collaboration of various stakeholders, including digital forensic investigators, cloud service providers, legal experts, and policymakers. Furthermore, it is essential to develop new approaches and technologies that can effectively address these challenges and ensure the admissibility of digital evidence in court.



\bibliographystyle{unsrt}
\bibliography{sample-bibliography}

\end{document}